\documentclass[epj]{svjour}
\usepackage{graphicx}
\usepackage{amsfonts}
\usepackage{amsmath}
\usepackage{bm}
\usepackage{color}
\usepackage{amssymb}
\usepackage{ulem}
\usepackage{times}
\usepackage{dcolumn}
\usepackage{cases}
\usepackage{txfonts}
\begin{document}

\title{Phase diagram and exotic spin-spin correlations of anisotropic Ising model on the Sierpi\'{n}ski gasket}

\author{Meng Wang\inst{1}, Shi-Ju Ran\inst{1}, Tao Liu\inst{1}, Yang Zhao\inst{2}, Qing-Rong Zheng\inst{1} and Gang Su\inst{1}
\thanks{\emph{Email: gsu@ucas.ac.cn} }%
}                     

\institute{Theoretical Condensed Matter Physics and Computational Materials Physics Laboratory, School of Physics, University of Chinese Academy of Sciences, P. O. Box 4588, Beijing 100049, China \and Department of Physics and Astronomy, University of Manitoba, Winnipeg, Canada R3T 2N2}
%
%
\abstract{The anisotropic antiferromagnetic Ising model on the fractal Sierpi\'{n}ski gasket is intensively studied, and a number of exotic properties are disclosed. The ground state phase diagram in the plane of magnetic field-interaction of the system is obtained. The thermodynamic properties of the three plateau phases are probed by exploring the temperature-dependence of magnetization, specific heat, susceptibility and spin-spin correlations. No phase transitions are observed in this model. In the absence of a magnetic field, the unusual temperature dependence of the spin correlation length is obtained with $0 \leq$J$_b/$J$_a<1$, and an interesting crossover behavior between different phases at J$_b/$J$_a=1$ is unveiled, whose dynamics can be described by the J$_b/$J$_a$-dependence of the specific heat, susceptibility and spin correlation functions. The exotic spin-spin correlation patterns that share the same special rotational symmetry as that of the Sierpi\'{n}ski gasket are obtained in both the $1/3$ plateau disordered phase and the $5/9$ plateau partially ordered ferrimagnetic phase. Moreover, a quantum scheme is formulated to study the thermodynamics of the fractal Sierpi\'{n}ski gasket with Heisenberg interactions. We find that the unusual temperature dependence of the correlation length remains intact in a small quantum fluctuation.
} 

\maketitle

\section{Introduction}

The classical models like Ising or Potts \cite{Potts} models often exhibit intriguing properties, especially when there exists geometrical frustration \cite{Frustrate} that may cause extensive ground-state degeneracy and nonzero residual entropy of the system at zero temperature \cite{ReEntro}. There are many interesting phenomena in these models that have been discovered but not yet been totally understood, for instance, the essence of the partial order \cite{PO} or the crossover between exotic phases, etc. Extensive works have been done on the Ising models in two- and three dimensions, such as the Ising models on the checkerboard lattice \cite{PO} and kagom\'{e} lattice \cite{liw}, as well as the Potts model on some irregular lattices \cite{POIrregular,zhaoy}, where the systems possess translational invariance. There were some early works on the Ising model on fractal lattices \cite{Fractal1,Fractal11,Fractal12,Fractal2,Fractal3,Fractal4,Fractal5,Fractal6}, which show the properties quite different from the models on one-, two- and three-dimensional lattices with translational invariance even with the simplest couplings. Recent reported on a new class of  highly frustrated two-dimensional magnetic materials $Cu_{9}X_{2}(cpa)_{6}\cdot xH_{2}O$(cpa=2-carboxypentonic acid, a derivative of ascorbic acid; X=F, Cl, Br) \cite{Material1,Material2,Material3,Material4,Material5,Material6} stimulate study of the spin-Heisenberg model on triangulated kagome lattice(TKL) \cite{Lattice1,Lattice2,Lattice3,Lattice4,Lattice5,Lattice6,Lattice7}. The TKL is actually an example of two-dimensional triangles-in-triangles (TIT) lattices \cite{TIT1,TIT2,TIT3}, which generally consist of smaller triangular entities embedded in either some or all triangular cells of 2D triangle-based lattices. Latest study on spin-$\frac{1}{2}$ Ising-Heisenberg model on two closely related TIT lattices reveal a profound relation between local quantum fluctions with overall magnetic behavior of this mixture classical-quantum spin model \cite{TIT1} and shed light the role of classical-quantum spin model on research the ground state of quantum Heisenberg model \cite{TIT2}. But for the Ising models on fractals, comprehensive studies on, e.g. spin-spin correlations where there is no translational invariance, are still needed and, the questions, such as how to describe the crossover behaviors in fractals, are yet to be explored.

The Sierpi\'{n}ski Ising model is so defined that the Ising spin is settled on each of vertices of the Sierpi\'{n}ski graph. The Hausdorff dimension $D_H$ of an Sierpi\'{n}ski graph satisfies
$D_H=\log($ $D_E+1) / \log(2)$, which is not an integer, where $D_E$ is its Euclidean dimension. Unlike other lattices such as square or kagom\'{e} lattices, a Sierpi\'{n}ski graph has no translational symmetries, but bears some special geometrical properties that can be seen from its recursive construction procedure. The $D_E=2$ Sierpi\'{n}ski graph, which is also named Sierpi\'{n}ski gasket (SG), is shown schematically in Fig. \ref{fig-SierpinskiRG} (a).

Some properties of the Ising model on Sierpi\'{n}ski graphs were noted. For example, the order of ramification on a SG is ``marginal'' \cite{Fractal12}, and by locating the fixed point of renormalization group (RG) equations, it was confirmed that there is no phase transition at any finite temperature \cite{Tc}.  With the nearest neighbor ferromagnetic couplings, the correlation length $\xi$ increases extremely fast with the inverse temperature $\beta = 1/T$ as \cite{Fractal12}
\begin{eqnarray}
 \xi \propto \exp [ \cfrac {\ln 2} {4} \exp (4J\beta) ],
\label{eq-CRL}
\end{eqnarray}
where $J$ is a positive constant (ferromagnetic coupling) and Boltzmann constant is taken as $k_B=1$. This unusual relation indicates that the correlation length $\xi$ is finite at any finite temperature. Since $\xi$ increases so fast with $\beta$,  upon lowering temperature, the correlation length $\xi$ can be larger than the size of any finite system, resulting in a long range order.

In this paper, the anisotropic antiferromagnetic Ising model on the SG is studied systematically in the thermodynamic limit. The ground state phase diagram in the plane of magnetic field ($h$)-interaction (J$_b$) is obtained, where three nontrivial phases, including the $1/3$ magnetization plateau ordered and disordered phases, and the $5/9$ plateau partially ordered ferrimagnetic phase, as well as five boundaries and two intersection points are identified. The $1/3$ plateau disordered phase and the $5/9$ plateau partially ordered ferrimagnetic phase are found to have extensive degeneracy. The thermodynamic properties of the system, including the temperature dependence of the uniform and subset magnetization, susceptibility, specific heat and spin correlation functions, are calculated. Some interesting results due to special geometrical properties like the fractional dimensionality and the absence of symmetries on usual periodic lattices are discovered. The nonexistence of the phase transition is testified. For $h=0$, the unusual relation between the correlation length and temperature is obtained for $0 \leq $J$_b/$J$_a<1$, and a nontrivial crossover behavior between different phases at J$_b/$J$_a=1$ is disclosed, whose dynamics can be described by the J$_b/$J$_a$-dependence of the specific heat, susceptibility and the spin correlation functions. Exotic spin-spin correlation patterns that have the same special rotational symmetry as that of the SG are observed in the $1/3$ plateau disordered phase. Moreover, we propose a scheme to study the thermodynamics of Sierpi\'{n}ski gasket with quantum interactions, and discover that the unusual relation of correlation length remains intact in a small quantum fluctuation.

This paper is organized as follows. In Sec. II, we introduce the anisotropic SG Ising model and give the exact formulation of the free energy. In Sec. III, the zero temperature phase diagram is identified by calculating the ground state uniform magnetization and residual entropy
\cite{Residual}. In Sec. IV, the temperature dependence of thermodynamic quantities, including the uniform and subset magnetization, susceptibility and specific heat, are studied. In Sec. V, the exotic spin-spin correlation functions of the three nontrivial plateau phases are disclosed. In Sec. VI, we generalize the scheme to study the thermodynamics of quantum models on the fractal Sierpi\'{n}ski gasket. Finally, a summary is given.

\section{Free energy of anisotropic Ising model on Sierpi\'{n}ski gasket}

\begin{figure}
\resizebox{0.46\textwidth}{!}{%
  \includegraphics{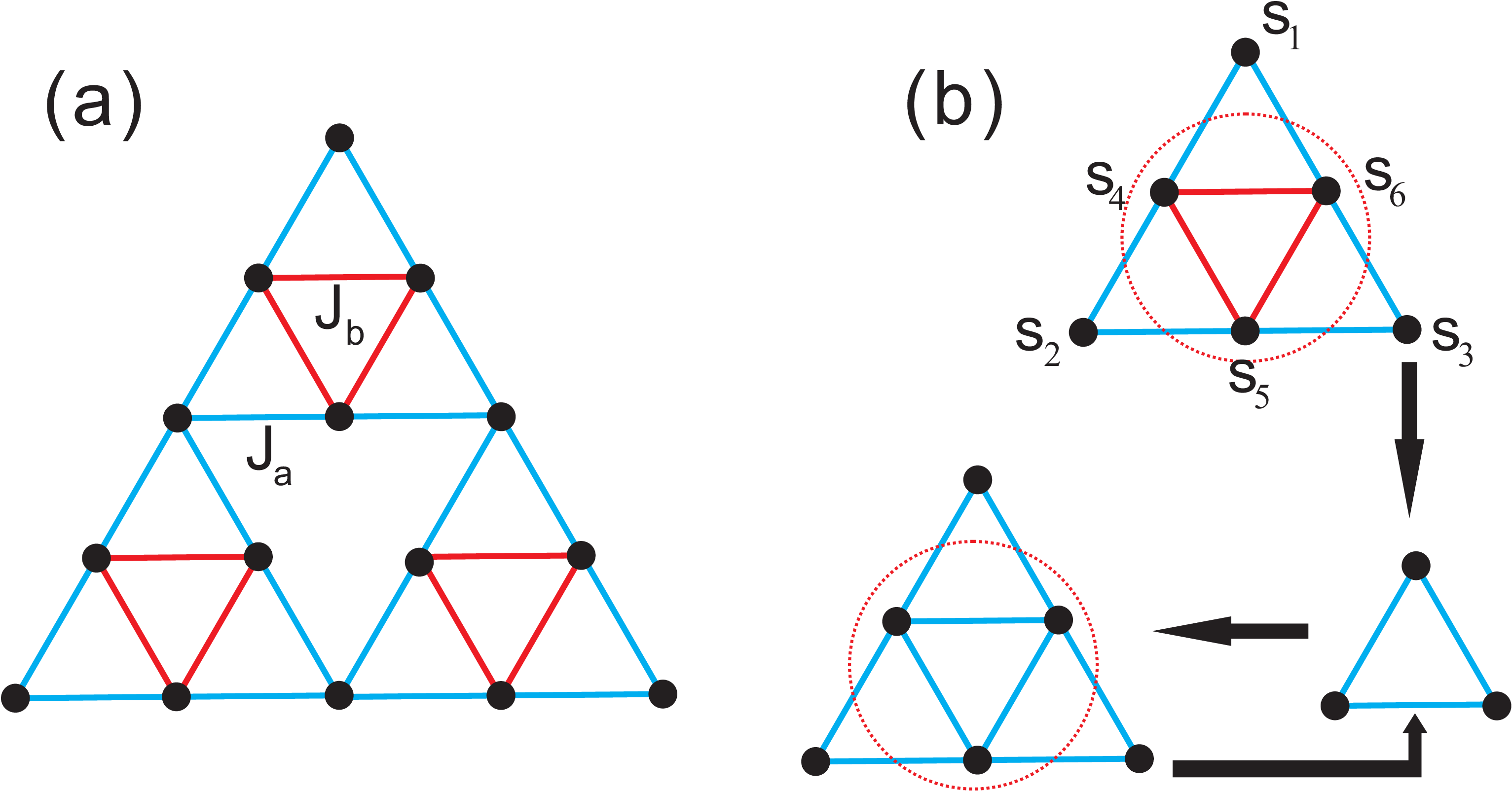}
}
\caption{(Color online) (a) Schematic depiction of Sierpi\'{n}ski gasket with the Euclidean dimension $D_E=2$ and the renormalization time $t=3$, where two kinds of couplings, J$_a$ (blue) and J$_b$ (red) are shown. (b) The real space renormalization of the SG Ising model. Each time after summing over the degrees of freedom inside the triangle (in the dash circle), the new triangle is obtained and used to construct the SG of trebled size in the next renormalization step.}
\label{fig-SierpinskiRG}
\end{figure}

Let us begin with discussing the exact formulation of free energy of the anisotropic antiferromagnetic Ising model on the SG with $D_E=2$. The Hamiltonian reads
\begin{eqnarray}
  H(\{ s \})=\sum_{ \langle ij \rangle } -J_{ij} s_i s_j-h(s_i+s_j)/4,
\label{eq-Hamiltonian}
\end{eqnarray}
where $\langle ij \rangle$ means the spins at vertices $i$ and $j$ that are nearest neighbors on the SG, each spin $s_i$ takes values of $\pm 1/2$, $h$ is the magnetic field and $J_{ij}$ is the negative coupling constant. The couplings are indicated in Fig. \ref{fig-SierpinskiRG} (a), where we presume $J_{ij}=J_a$ for the blue bonds and $J_{ij}=J_b$ for the red bonds. We denote the SG that contains $N^{(t)} = (3+3^t)/2$ spins as $\Delta^{(t)}$ and its density matrix as $\rho^{(t)} (\{ s \}) = \exp [-\beta H(\{ s \})]$, where $t$ is the renormalization step. Then, the partition function $Z$ can be expressed as $Z^{(t)}=\sum_{\{ s \}} \rho^{(t)} (\{ s \})$.

We begin with $\rho^{(2)}$, which is the density matrix of the model with only six Ising spins in $\Delta^{(2)}$ [Fig. \ref{fig-SierpinskiRG} (b)] as
\begin{eqnarray}
  \rho^{(2)} = \exp \{ -\beta \sum_{\langle ij \rangle\in \Delta^{(2)}} [-J_{ij} s_i s_j-h(s_i+s_j)/4]\}.
\label{eq-InitialT}
\end{eqnarray}
The density matrix $\rho^{(2)}$ for $\Delta^{(2)}$ can be represented by
\begin{equation}
\begin{split}
  \rho^{(2)}_{s_1s_2s_3s_4s_5s_6}=
exp \{ -\beta [ -J_a\{s_1s_4 +s_1s_6 +s_4s_2
  +s_2s_5 +\\s_5s_3+s_3s_6\} - J_b \{s_4s_5+ s_5s_6+ s_6s_4\}
  -h(s_1+s_2+ s_3)/2 - \\h(s_4+s_5+s_6) ] \}.
\label{eq-InitialT}
\end{split}
\end{equation}
Now, we introduce the ``corner density matrix'' $f^{(t)}$ by summing over the degrees of freedom of $\rho^{(t)}$ except for the spins at the three vertices of the largest triangle in the $\Delta^{(t)}$. For example, we have $f^{(2)}_{s_1s_2s_3} = \sum_{s_4s_5s_6} \rho^{(2)}_{s_1s_2s_3 s_4s_5s_6}$, where $s_1$, $s_2$ and $s_3$ are the three spins at the three vertices of the largest triangle in the $\Delta^{(2)}$. The recursive relation of the real space renormalization of the corner density matrix can be given by decimating the degrees of freedom inside $\Delta^{(t)}$ [Fig. \ref{fig-SierpinskiRG} (b)] as
\begin{eqnarray}
 f^{(t)}_{s_as_bs_c} = \sum_{s_a's_b's_c'} f^{(t-1)}_{s_as_a's_c'} f^{(t-1)}_{s_a's_bs_b'} f^{(t-1)}_{s_c's_b's_c} (f^{(t-1)}_{1,1,1})^{-3},
\label{eq-RecursiveTensor}
\end{eqnarray}
where $s_a$, $s_b$ and $s_c$ denote the spins at the corner of the largest triangle in $\Delta^{(t)}$. As a matter of fact, Eq. (\ref{eq-RecursiveTensor}) can be considered as tensor contractions. This real space renormalization method is equivalent to a tensor renormalization group approach \cite{TRG} on a fractal lattice. The factor $(f^{(t-1)}_{1,1,1})^{-3}$ is introduced to avoid divergence.

The partition function of the Ising model on $\Delta^{(t)}$ with $N^{(t)}=(3+3^t)/2$ spins becomes
\begin{equation}
Z^{(t)} = k \sum_{s_as_bs_c} f^{(t)}_{s_as_bs_c},
\end{equation}
with $k=\prod_{\tau=2}^{t-1} (f^{(\tau)}_{1,1,1})^{3^{t-\tau}}$. So, the free energy per site $\mathcal{F}^{(t)}$ can be obtained by
\begin{eqnarray}
\begin{split}
 \mathcal{F}^{(t)}= - \cfrac {\ln Z^{(t)}} {N^{(t)}\beta} =& -\cfrac{3^{t}} {N^{(t)} \beta}[ \sum_{\tau=2}^{t-1} (3^{-\tau} \ln f^{(\tau)}_{1,1,1})+\\&3^{-t} \ln (\sum_{s_as_bs_c} f^{(t)}_{s_as_bs_c}) ].
\label{eq-FreeEnergy}
\end{split}
\end{eqnarray}
When $t\gg1$, one can take $3^{t}/N^{(t)} \simeq 2$, and $3^{-t} \ln (\sum_{s_as_bs_c} f^{(t)}_{s_as_bs_c}) \simeq 0$ because $\ln (\sum_{s_as_bs_c} f^{(t)}_{s_as_bs_c})$ is finite.

The thermodynamic quantities such as the energy per site, specific heat, entropy, uniform magnetization and susceptibility can be calculated through the differentiation of the free energy given in Eq.(\ref{eq-FreeEnergy}).

Moreover, the magnetization of the single spin $\langle s_a \rangle$ and the spin-spin correlation function $\langle s_a s_b \rangle$, defined by
\begin{eqnarray}
\langle s_a \rangle&=&\sum_{\{ s \}} \{\exp[ -\beta E(\{s\}) ] s_a\}/Z,
\label{eq-Mag}
\\
\langle s_a s_b \rangle &=& \sum_{\{ s \}} \{\exp[ -\beta E(\{s\}) ] s_a s_b\}/Z,
\label{eq-CRL}
\end{eqnarray}
can be obtained by performing the calculations similar to Eq. (\ref{eq-RecursiveTensor}). We take $\langle s_1 \rangle$ as an example. Considering that the only difference between the equations for $Z$ and $\langle s_1 \rangle$ is the contraction that involves $s_1$, we can introduce the impurity as $\tilde{f}^{(2)}_{s_1s_2s_3} = f^{(2)}_{s_1s_2s_3}(s_1-3/2)$, and then, its renormalization can be written as
\begin{eqnarray}
\tilde{f}^{(t+1)}_{s_as_bs_c} = \sum_{s_a's_b's_c'} \tilde{f}^{(t)}_{s_as_a's_c'} f^{(t)}_{s_a's_bs_b'} f^{(t)}_{s_c's_b's_c} (f^{(t)}_{1,1,1})^{-3},
\label{eq-Impurity}
\end{eqnarray}
where each $f^{(t)}$ satisfies Eq. (\ref{eq-RecursiveTensor}). Compared with Eq. (\ref{eq-RecursiveTensor}), one of the three $f^{(t)}$'s in Eq. (\ref{eq-Impurity}) is replaced by the impurity, and which tensor should be replaced depends on the position of $S_1$ in the SG. In the end we have $\langle s_1 \rangle = \sum_{s_as_bs_c} \tilde{f}^{(t)}_{s_as_bs_c} / \sum_{s_ds_gs_h} f^{(t)}_{s_ds_gs_h}$. The calculation of $\langle s_a s_b \rangle$ is similar. It should be mentioned that the whole formulation above can be readily extended to the N-state Potts model \cite{Potts,zhaoy}.

The above formulation can be further simplified in the absence of a magnetic field ($h=0$), and the fixed point of the renormalization can be discussed. Under this circumstance, each tensor $f^{(t)}$ only contains two inequivalent components denoted as $x^{(t)} = f^{(t)}_{1,1,1} = f^{(t)}_{2,2,2}$ and $y^{(t)} = f^{(t)}_{2,1,1} = f^{(t)}_{1,2,1} = f^{(t)}_{1,1,2} = f^{(t)}_{2,2,1} = f^{(t)}_{2,1,2} = f^{(t)}_{1,2,2}$. According to Eq. (\ref{eq-RecursiveTensor}), we can readily get the recursive relation for $z^{(t)}=y^{(t)} / x^{(t)}$ as \cite{Recursive}
\begin{eqnarray}
z^{(t+1)}= \cfrac { 3[z^{(t)}]^3 + 4[z^{(t)}]^2 + z^{(t)}} {4[z^{(t)}]^3 + 3[z^{(t)}]^2 +1},
\label{eq-Recursivez}
\end{eqnarray}
with $z^{(1)} = f^{(1)}_{1,1,2} / f^{(1)}_{1,1,1}$. Consequently, the free energy becomes
\begin{eqnarray}
\begin{split}
\mathcal{F}^{(t)}=- 2\{\sum_{\tau=2}^{t-1} [3^{-\tau} \ln(4[z^{(\tau)}]^3 + 3[z^{(\tau)}]^2 +1)]+ \\3^{-1} \ln f^{(2)}_{1,1,1} + 3^{-t} \ln8 \} /\beta .
\label{eq-FreeEnergyZ}
\end{split}
\end{eqnarray}
Eq. (\ref{eq-Recursivez}) has two fixed points: $z=0$ (unstable) and $z=1$ (stable). By considering that $f^{(t)}_{s_as_bs_c}$ represents the probability distribution of an effective Ising model of the three spins, one can see that $z=0$ implies the effective temperature $T^{eff}=0$ (infinite inverse temperature), and $z=1$ implies $T = \infty$ when all configurations of spins share the same probability, indicating that no phase transition happens at any finite temperature. This is consistent with the previous result in e.g. \cite{Tc}.

\section{Phase diagram at zero temperature}

\begin{figure}
\resizebox{0.45\textwidth}{!}{%
  \includegraphics{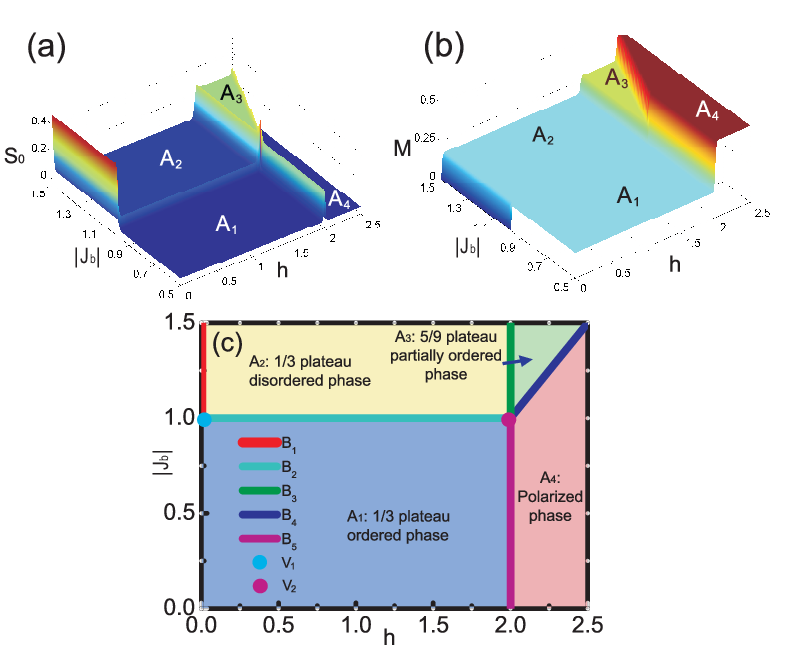}
}
\caption{(Color online) (a) The residual entropy $S_0$ and (b) the uniform magnetization per site $M$ against magnetic field $h$ and coupling $|J_b|$. (c) The zero temperature phase diagram of the antiferromagnetic SG Ising model. There are four phases, including A$_1$ phase: the $1/3$ magnetization plateau phase with ferrimagnetic order and $S_0=0$; A$_2$ phase: the $1/3$ plateau disordered phase with $S_0=0.0257$; A$_3$: the $5/9$ plateau partially ordered ferrimagnetic phase with $S_0 = 2/9\ln3 = 0.2441$; and A$_4$ phase: the polarized phase. The residual entropy at the phase boundaries: $S_0(B_1)=0.4752$, $S_0(B_2)=0.0770$, $S_0(B_3)=0.2738$, $S_0(B_4)=0.3081$, and $S_0(B_5)=0.2311$; and at the two intersection points $S_0(V_1)=0.4930$, and $S_0(V_2)=0.3843$. Here we take $\beta=200$.}
\label{fig-Ground}
\end{figure}

By calculating the residual entropy $S_0$ and the uniform magnetization $M$ in a magnetic field $h$ at sufficiently large inverse temperature \cite{RGT}, we have found four phases that are marked by A$_1$, A$_2$, A$_3$ and A$_4$ (Fig. \ref{fig-Ground}). For $0\leq h<2$ and $0 \leq |J_b|<1$ (we take $J_a=-1$ as energy scale and $J_b<0$ for the antiferromagnetic coupling), the system is in the A$_1$ phase, where the residual entropy $S_0=0$ and the uniform magnetization $M=1/6$. After grouping the spins into four subsets denoted as P$_1$, P$_2$, P$_3$ and P$_4$ [Fig. \ref{fig-S0_Mag} (a)], we calculated the zero temperature subset magnetization and obtained $M_{1,3}=-1/2$ and $M_{2,4}=1/2$ with $M_{i}$ ($i=1,2,3,4$) the subset magnetization of $P_i$. It can be seen that the system is in the ferrimagnetic order, where $2/3$ of the spins that point up and belong to P$_2$ and P$_4$ subsets, and $1/3$ of them that point down and belong to P$_1$ and P$_3$ subsets, resulting in a $1/3$ magnetization plateau phase.

\begin{figure}
\resizebox{0.45\textwidth}{!}{%
  \includegraphics{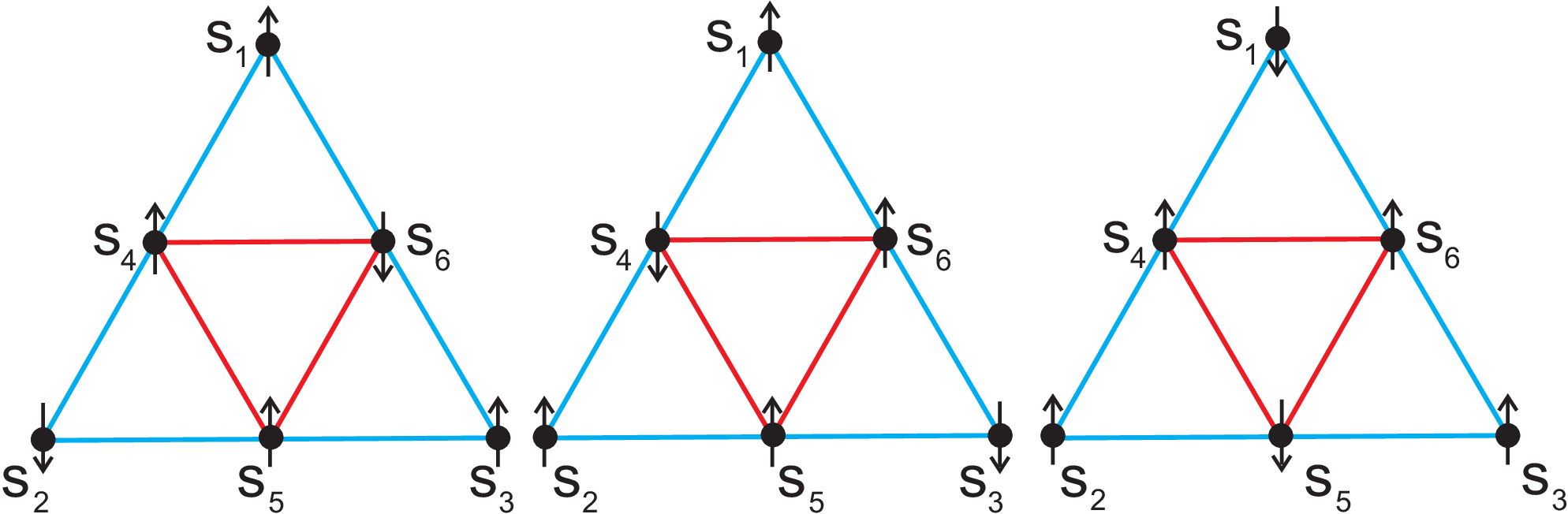}
}
\caption{(Color online) The spin configurations of elementary cell of the macroscopically degenerate $\frac{1}{3}$-plateau state $A_2$.}
\label{fig-Spin Configuration}
\end{figure}

Keeping $0<h<2$ and increasing the coupling from $|J_b|<1$ to $|J_b|>1$, we observed that the system enters into the A$_2$ phase (e.g. $|J_b|=1.2$, $h=1$) with the same magnetization $M=1/6$, but the residual entropy is nonzero, $S_0=0.0257$, showing that this phase is in a disordered state with extensive ground state degeneracy. The spin configurations of elementary cell of the macroscopically degenerate $\frac{1}{3}$-plateau state $A_2$ shows in Fig. \ref{fig-Spin Configuration}. The subset magnetization in A$_2$ phase is $M_{1,4}=0$ and $M_{2,3}=1/4$, which is totally different from those in A$_1$ phase. It can be seen that the residual entropy in A$_2$ phase is associated with all spins, so A$_2$ phase is of a disordered $1/3$ plateau phase. This fact indicates that by increasing $|J_b|$, the ferrimagnetic order in A$_1$ phase is ``melted'' in A$_2$ phase by the classical frustration whereas the uniform magnetization is unchanged.

Noting that there are also exist $1/3$ plateau in both the spin-$\frac{1}{2}$ Ising-Heisenberg model and the pure quantum spin-$\frac{1}{2}$ Heisienberg model on two closely related TIT lattices at zero temperature in Ref.\cite{TIT2}.We true believe that the ¡°triangles-in triangles¡± structure is responsible for the presence of plateau. This is illuminating to explore the effects of geometric frustration. For $h>2$ and $|J_b|>h-1$, the system is in the A$_3$ phase with $M=5/18$ and $S_0=(2\ln 3)/9$, and the subset magnetization satisfies $M_{1,3}=1/2$ and $M_{2,4}=1/6$. The spins belonging to P$_1$ and P$_3$ are totally polarized, and only the spins belonging to P$_2$ and P$_4$ make contributions to the residual entropy. Thus, this phase is a partially ordered $5/9$ plateau ferrimagnetic phase.

For $h>2$ and $0\leq |J_b|<h-1$, the system is in the A$_4$ phase with $M=1/2$ and $S_0=0$, which is nothing but a polarized state.

For $h=0$, when $0\leq |J_b| <1$ the system bears the ferrimagnetic order with spontaneous magnetization $M=1/6$, but on the contrary, when $|J_b| \geq 1$ [the boundary $B_1$ in Fig. \ref{fig-Ground} (c)], we found the subset magnetization becomes zero, which indicates that $|J_b|=1$ may be a crossover line. Such a crossover behavior will be re-examined by studying the specific heat, susceptibility and spin-spin correlation functions in Secs. IV and V. In addition, it appears that the magnetization in A$_2$ phase cannot appear spontaneously but can only be induced by a magnetic field.

It should be point out that in our calculation, we find that the boundaries of phases are continuous and there is no singularity at any finite temperatures. However, when we reduce the temperature, we find that the magnetization approaches to a step function,which implies a first-order transition driven either by the external magnetic field or the interaction term $|J_b|$.

We also calculated the residual entropy $S_0$ at each of phase boundaries [Fig. \ref{fig-Ground} (c)], and disclosed that $S_0$ are higher than those of the surrounding phases. At boundary $B_1$ with $h=0$ and $|J_b|>1$ which is the left boundary of A$_2$ phase, we got $S_0(B_1)=0.4752$; at boundary $B_2$ between A$_1$ and A$_2$ phases with $0< h <2$ and $|J_b|=1$, we found $S_0(B_2)=0.0770$; at boundary $B_3$ between A$_2$ and A$_3$ phases with $h=2$ and $|J_b|>1$, we had $S_0(B_3)=0.2738$; at boundary $B_4$ between A$_3$ and A$_4$ phases with $h=|J_b|+1$ and $|J_b|>1$, we obtained $S_0(B_4)=0.3081$; at boundary $B_5$ between A$_1$ and A$_4$ phases with $h=2$ and $0<|J_b|<1$, we arrived at $S_0(B_4)=0.2311$. In particular, we observed that $S_0$ at the two intersection points of the boundaries are larger than those at the related boundaries, namely $S_0(V_1)=0.4930$ at the intersection point ($V_1$) between $B_1$ and $B_2$, and $S_0(V_2)=0.3843$ at the intersection point ($V_2$) between $B_2$, $B_3$, $B_4$ and $B_5$.

\section{Thermodynamic Properties}

In this section, we shall study the temperature dependence of the thermodynamic quantities including the uniform and subset magnetization, specific heat, susceptibility, where the nonexistence of phase transitions is testified, and the thermodynamic properties of the system will be explored.

\begin{figure}[tbp]
\resizebox{0.45\textwidth}{!}{%
  \includegraphics{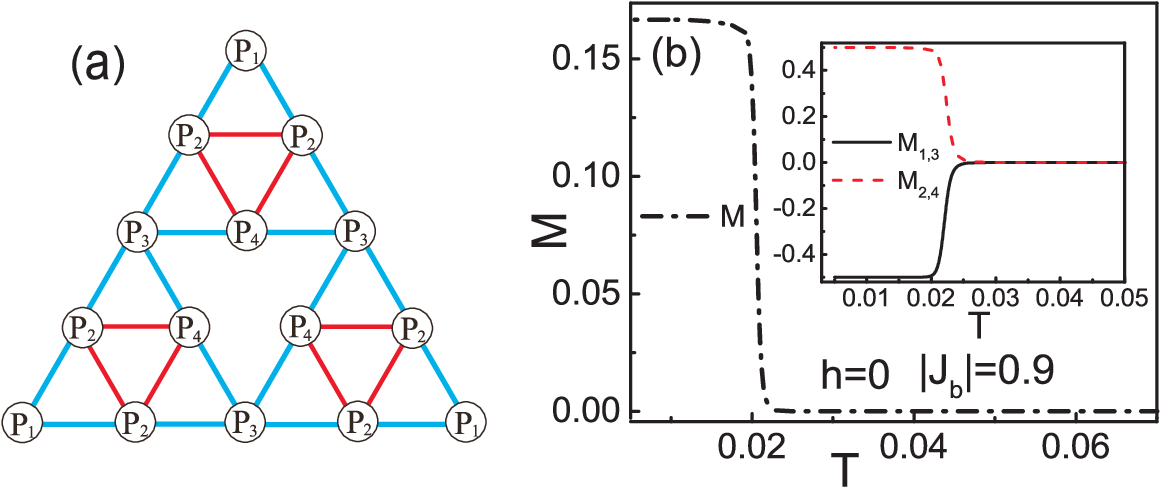}
}
\caption{(Color online) (a) The spins are divided into four subsets denoted as P$_1$, P$_2$, P$_3$ and P$_4$. (b) The $T$-dependence of the uniform magnetization per site at $h=0$ and $|J_b|=0.9$, where a $1/3$ magnetization plateau appears at low temperatures. The inset shows the subset magnetization against $T$.}
\label{fig-S0_Mag}
\end{figure}

Let us first look at the uniform magnetization [Fig. \ref{fig-S0_Mag} (b)] and subset magnetization [the inset of Fig. \ref{fig-S0_Mag} (b)] versus $T$ in A$_1$ phase by taking $|J_b|=0.9$ and $h=0$. One may see that a $1/3$ magnetization plateau appears at low temperature, and the magnetization becomes zero smoothly from nonzero at around $T=0.02$, which indicates that the system is crossing from the low-temperature ferrimagnetic phase to the high temperature disordered paramagnetic phase.

\begin{figure}[tbp]
\resizebox{0.45\textwidth}{!}{%
  \includegraphics{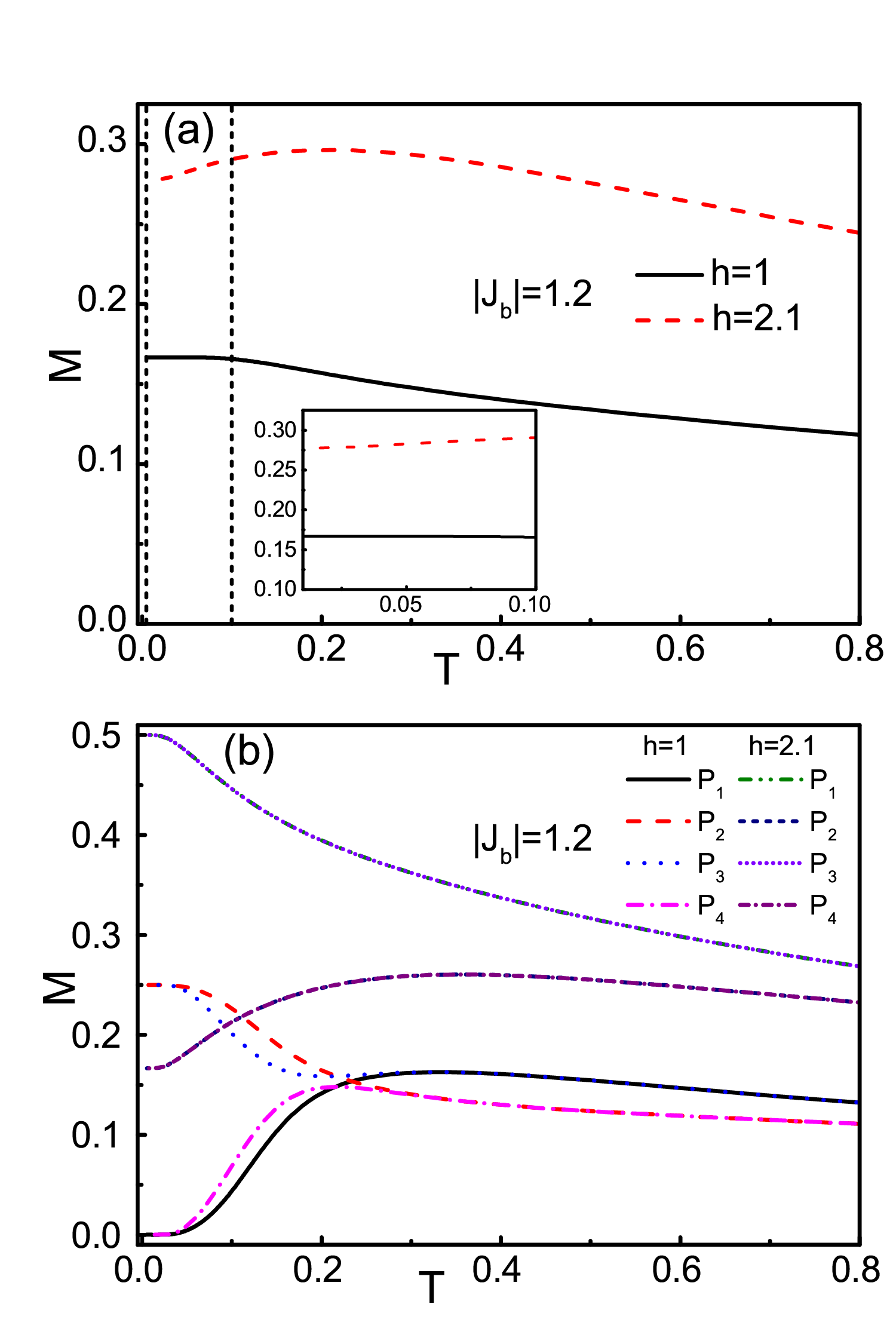}
}
\caption{(Color online) The $T$-dependence of (a) the uniform magnetization and (b) the subset magnetization in A$_2$ ($h=1$) and A$_3$ ($h=2.1$) phases. Here we take $|J_b|=1.2$.}
\label{fig-MagSingle}
\end{figure}

The uniform and subset magnetization in A$_2$ and A$_3$ phases [Fig. \ref{fig-MagSingle}] also have the similar behaviors, showing that the system is crossing into the high-temperature phases. In A$_2$ phase, the spins belonging to P$_1$ and P$_4$ exhibit similar behaviors, where the subset magnetization first rises from zero to its maximum at about $T=0.25$ and then decreases to zero with increasing $T$, while the spins at P$_2$ and P$_3$ sites behave similarly, where the $1/4$ saturation magnetization decreases to zero with increasing $T$. In A$_3$ phase, the situation is somehow different, where the spins at P$_1$ and P$_3$ sites are gradually decreases from polarized to zero , and the subset magnetization of P$_2$ and P$_4$ first converges to $1/6$ and then reaches the maximum as $T$ increases.

\begin{figure}[tbp]
\resizebox{0.45\textwidth}{!}{%
  \includegraphics{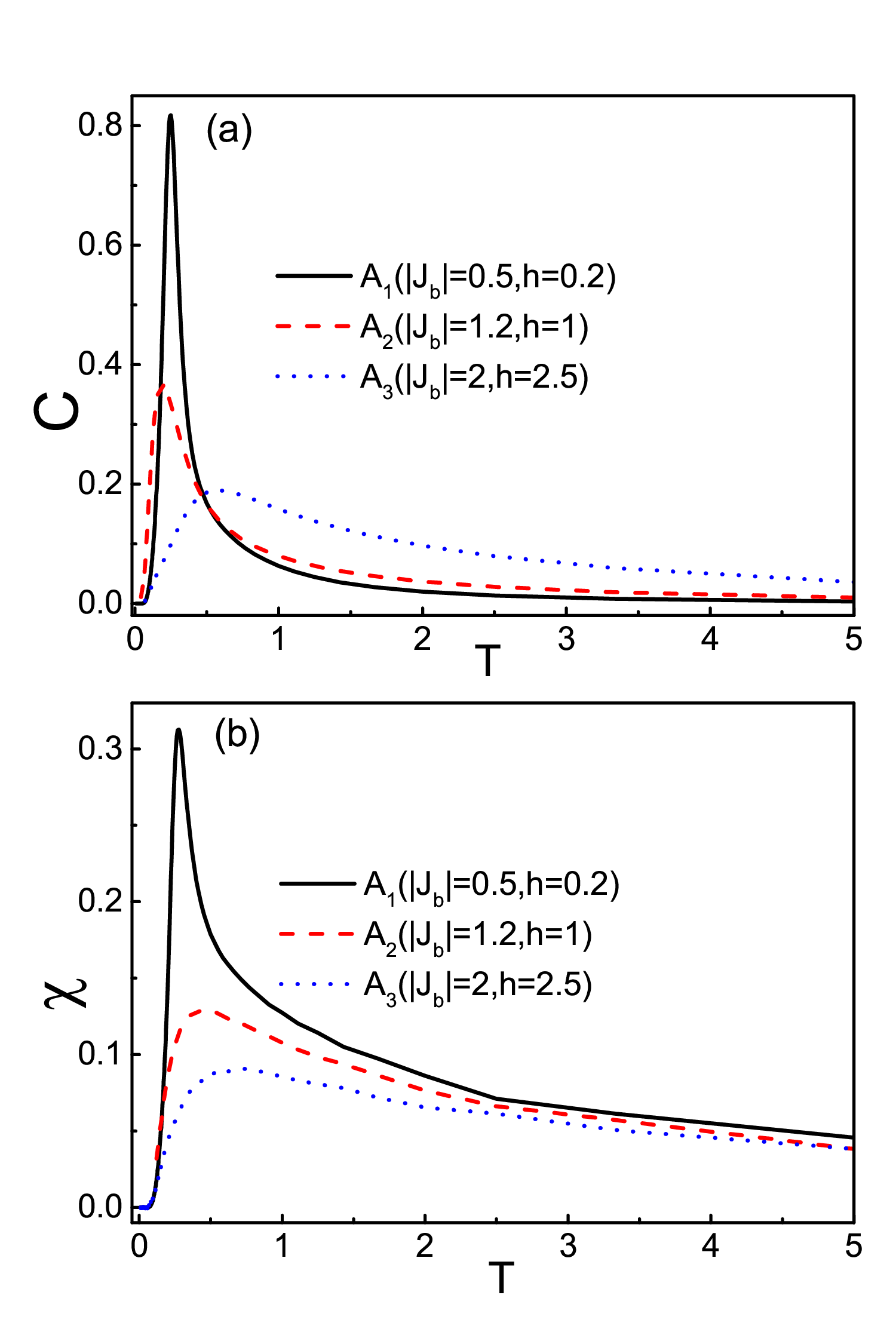}
}
\caption{(Color online) The $T$-dependence of (a) the specific heat $C$ and (b) the susceptibility $\chi$ at $|J_b|=0.5$ and $h=0.2$ in A$_1$ phase, $|J_b|=1.2$ and $h=1$ in A$_2$ phase, and $|J_b|=2$ and $h=2.5$ in A$_3$ phase.}
\label{fig-KaiC}
\end{figure}

Fig. \ref{fig-KaiC} shows the specific heat $C$ and the susceptibility $\chi$ versus $T$ in three nontrivial A$_1$, A$_2$ and A$_3$ phases. It indicates that the system enters the high-temperature phases through crossover instead of phase transitions (even though the peak in A$_1$ phase is relatively shaper, it is still round and continuous). One may notice that the temperature at which the peak of susceptibility appears is slightly higher than that of the peak of specific heat, where the fluctuation of energy reaches the maximum.

\begin{figure}[tbp]
\resizebox{0.45\textwidth}{!}{%
  \includegraphics{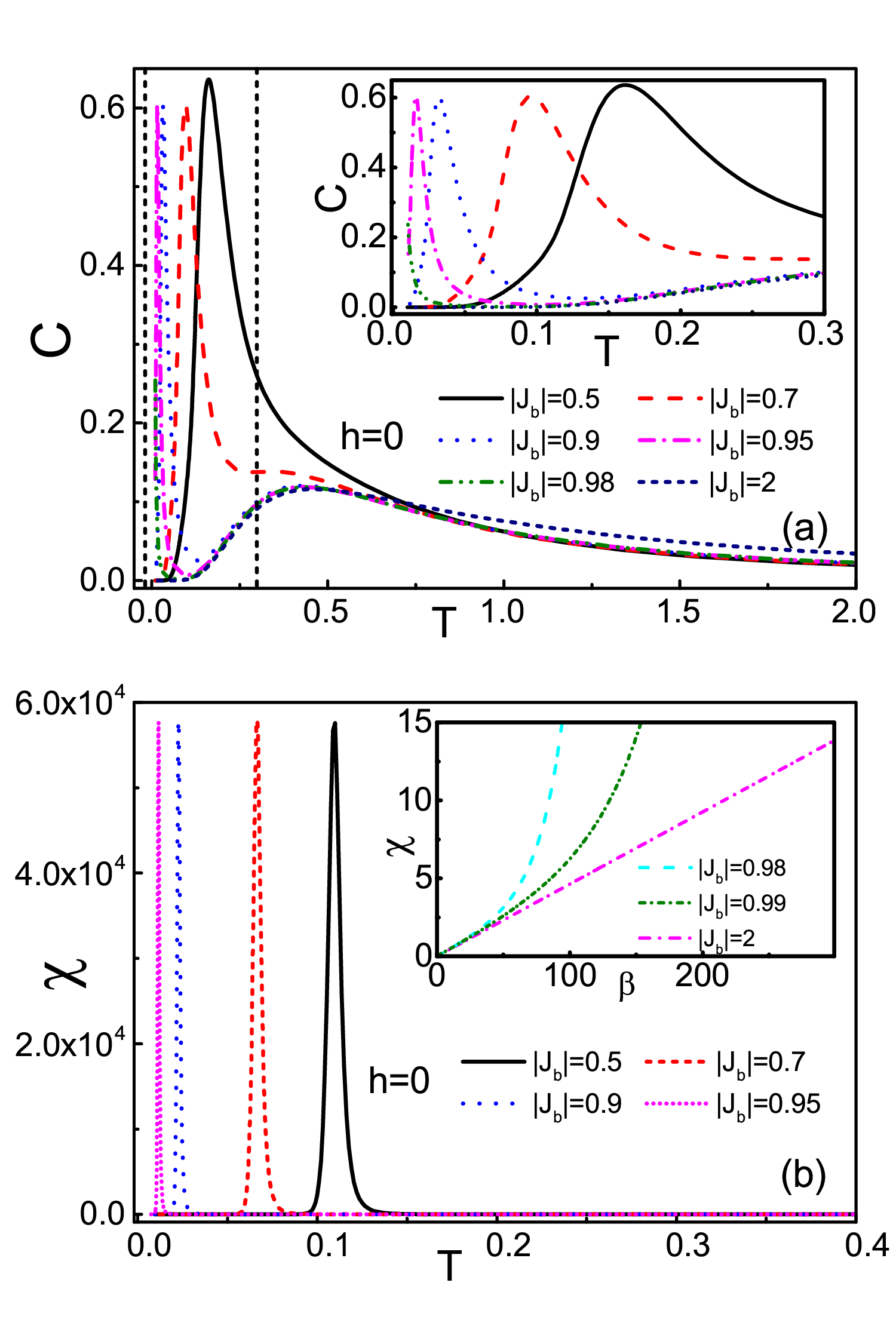}
}
\caption{(Color online) The $T$-dependence of (a) the specific heat $C$ and (b) susceptibility $\chi$ at $h=0$ for different $|J_b|$. In (a), the specific heat has two peaks, where the position of high-temperature peak is independent of $|J_b|$, while the position of the low-temperature peak, which lies in the finitely correlated region shown in Fig. \ref{fig-CrossoverCRL}, moves to infinitesimal $T$ as $|J_b| \rightarrow 1$. In (b), there is only one peak for the susceptibility, whose position also approaches infinitesimal as $|J_b| \rightarrow 1$. It can be seen in the inset of (b) that $\chi$ gradually becomes proportional to $\beta=\frac{1}{T}$ as $|J_b| \rightarrow 1$ ($|J_b|=0.98$, $0.99$). When $|J_b| \geq 1$, $\chi$ bears a linear relation with $\beta=\frac{1}{T}$.}
\label{fig-SpecificA1}
\end{figure}
In Sec. III, we have identified $|J_b|=1$ as a crossover line for $h=0$. Here, we studied this crossover behavior by calculating the specific heat $C$ and susceptibility $\chi$ for different $|J_b|$, as shown in Fig. \ref{fig-SpecificA1}. For $|J_b| \geqslant 1$ (including the point $|J_b|=1$), there is only one peak of $C$ at high temperature $T \simeq 0.45$ which is independent of $|J_b|$. With gradually decreasing $|J_b|$ to $|J_b|<1$, another peak moves from the low-temperature side to the high-temperature side, while the position of the high-temperature peak remains unchanged. This double-peak structure indicates a mixture of two kinds of excitations corresponding to the two phases. For $|J_b|$ smaller than $0.5$, the two peaks merge into one, and the low-temperature peak stands dominant. For the susceptibility $\chi$, there is only one peak, whose position moves to infinitesimal as $|J_b| \rightarrow 1$. This peak, although it looks apparently divergent, is still finite and broad. Meanwhile, as shown in Fig. \ref{fig-SpecificA1} (b), the relation between $\chi$ and $\beta$ becomes more and more linear as $|J_b| \rightarrow 1$. When $|J_b| \geq 1$, $\chi$ bears a linear relation with $\beta$. In the following section, we will show for different $|J_b|$, the positions of both the low-temperature peaks of the specific heat and the susceptibility locate within the finite correlated region.

\section{Exotic spin-spin correlations}

Now we study the spin-spin correlation function $\zeta = \langle s_a s_b \rangle$ [Eq. (\ref{eq-CRL})] in the three non-trivial phases. The A$_1$ phase is manifested to be with a Ferrimagnetic order, and at $h=0$ it has an unusual long range correlation (ULRC) which is similar to the behavior characterized by Eq. (\ref{eq-CRL}) and the long range correlation is in the usual sense in the presence of $h$. We also discovered that the ULRC corresponds to an intermediate region which separates the $|J_b| - \beta$ diagram into two regions, the fully correlated region with the correlation length $\xi = \infty$ and the non-correlated region with $\xi < 2$. In the ULRC region where the system is finitely correlated, $\xi$ increases ``super-exponentially'' with $\beta$ until the system enters into the fully correlated region. When $|J_b| \rightarrow 1$, $\beta_0$ and $\beta_1$, the upper and lower boundaries of the ULRC region, both approach infinite, and so consequently, the system can only stay in the non-correlated region at finite temperature for $|J_b| \geq 1$.

The correlations in the A$_2$ and A$_3$ phases are more complicated, where exotic correlation patterns that have same special rotation symmetry as that of the SG are disclosed below. The A$_4$ phase is a fully polarized phase and will not be discussed here.

\subsection{Unusual long range correlation for $0 \leq |J_b| <1$ and $h=0$}

\begin{figure}[tbp]
\resizebox{0.45\textwidth}{!}{%
  \includegraphics{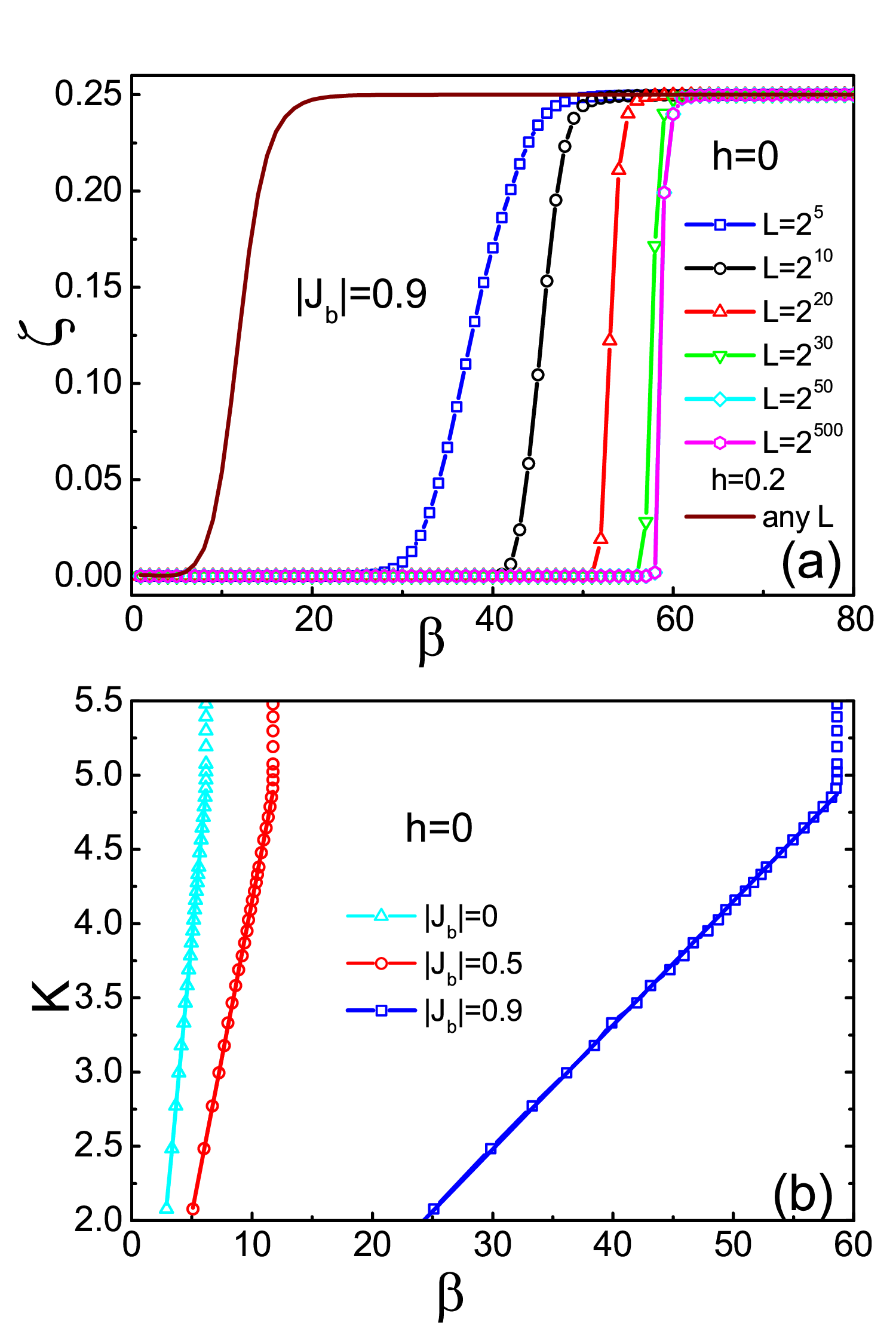}
}
\caption{(Color online) (a) The $\beta$-dependence of the correlation function $\zeta$ for different distances $L$ at $|J_b|=0.9$. (b) The $\beta$-dependence of $ K = \ln[4 \ln\xi /\ln2] /4$ as well as their linear fittings. The regression coefficient of each fitting is larger than $0.999$. }
\label{fig-CRLAF}
\end{figure}

The spin-spin correlation functions $\zeta$ versus $\beta$ for $0 \leq |J_b| <1$ and $h=0$ for different distances $L$ between two spins are studied. Fig. \ref{fig-CRLAF} (a) shows the results at $|J_b|=0.9$. By defining the distance where the system bears the correlation $\zeta = 1/(4e)$ as the correlation length $\xi$, we determine the $\beta$-dependence of $\xi$. We discovered that in this ferrimagnetic phase for $\beta$ smaller than a certain value denoted as $\beta_1$, the correlation length shares the similar unusual form as Eq. (\ref{eq-CRL}), where $\xi$ increases extremely fast when $\beta$ increases. We show the high linearity of the relation between $\beta$ and $K=\ln (4 \ln \xi /\ln 2)$, i.e. $K = k_1 \beta + k_2$ [Fig. \ref{fig-CRLAF} (b)], which gives rise to the temperature dependence of the correlation length having the simple form of
\begin{eqnarray}
\xi= \exp[\cfrac {\ln 2} {4} c_2 \exp( c_1 \beta )],
\label{eq-xi}
\end{eqnarray}
where the coefficients $c_1$ and $c_2$ are determined by $c_1=k_1$ and $c_2=e^{k_2}$. The values of $k_1$ and $k_2$ for several $|J_b|$'s are shown in Table \ref{tab-linear} for useful information. Also, it should be noted that the regression coefficient $R^2$ of each linear fitting is larger than $0.999$. The correlation function is independent of the distance $L$ in the presence of $h$.

\begin{table}[tbp]
\caption{ The coefficients $k_1$ and $k_2$ for different $|J_b|$}
\begin{tabular*}{8cm}{@{\extracolsep{\fill}}lcccc}
\hline\hline
$|J_b|$ & 0 & 0.2 & 0.5 & 0.9 \\ \hline
$k_1$ & 0.8547 & 0.6907 & 0.4206 & 0.0835 \\
$k_2$ &-0.375 & -0.2462 & -0.054 & -0.0251 \\
$R^2$ &0.9995&0.9994&0.9996&0.9995
 \\ \hline\hline
\label{tab-linear}
\end{tabular*}
\end{table}

\begin{figure}[tbp]
\resizebox{0.42\textwidth}{!}{%
  \includegraphics{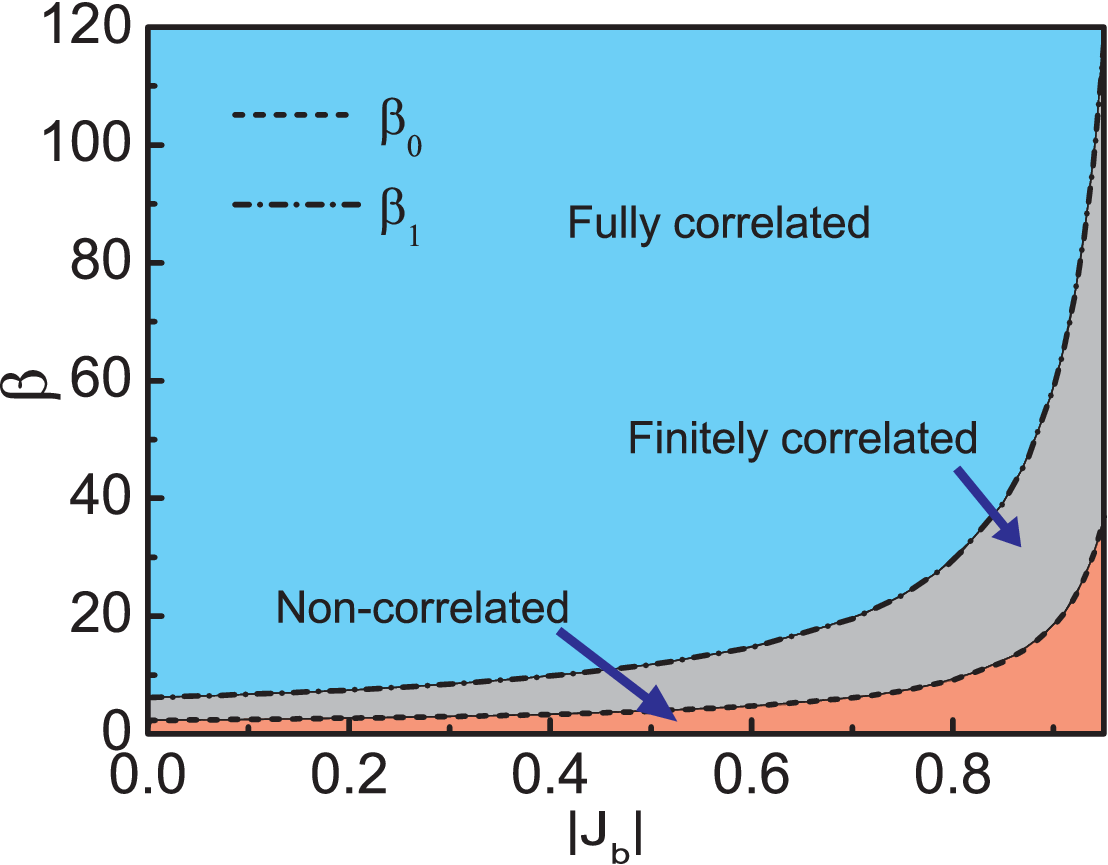}
}
\caption{(Color online) The $|J_b|$-dependence of $\beta_0$ and $\beta_1$ which separate the diagram into three subregions. In the subregion beyond the $\beta_1$ line, the system is fully correlated with the correlation length $\xi=\infty$; in the subregion below the $\beta_0$ line, the system is non-correlated with $\xi<2$; and in the subregion between the two lines where the system is finitely correlated, the correlation length $\xi$ obeys Eq. (\ref{eq-xi}) in which $\xi$ increases ``super-exponentially'' with $\beta$. }
\label{fig-CrossoverCRL}
\end{figure}

Also, by locating the inverse temperature $\beta_0$ where the correlation length is $\xi=2$, we can find where the system begins to be correlated. Fig. \ref{fig-CrossoverCRL} shows the $|J_b|$-dependence of $\beta_0$ and $\beta_1$, where the whole $\beta$-$|J_b|$ region is separated into three subregions. In the subregion beyond the $\beta_1$ line, the system is fully correlated with $\xi=\infty$; in the subregion below the $\beta_0$ line, the system is non-correlated with $\xi<2$; in the subregion between two lines where the system is finitely correlated, the correlation length $\xi$ obeys Eq. (\ref{eq-xi}) in which $\xi$ increases ``super-exponentially'' with $\beta$.

From Fig. \ref{fig-CrossoverCRL}, the behavior of $\beta_0$ and $\beta_1$ when $|J_b|\rightarrow 0$ as well as $|J_b| \rightarrow 1$ can be obtained. With $|J_b|\rightarrow 0$, the two curves manifest a linear relation with $|J_b|$. With $|J_b| \rightarrow 1$ from the $|J_b|<1$ side, both $\beta_0$ and $\beta_1$ go divergent with the speed faster than the exponential divergence. Such a behavior of $\beta_0$ can provide a clear picture of what happens during the crossover at $|J_b|=1$ in the absence of a magnetic field. With $|J_b| \rightarrow 1$, the system needs a much lower temperature crossover to the finitely correlated region. At $|J_b|=1$, as $\beta_0$ becomes infinite, the system is forbidden to crossover to the finitely correlated region at any finite temperature. Thus, $|J_b|=1$ is a crossover line. Meanwhile, this is coincident with the results of the low-temperature peaks of $C$ and $\chi$ shown in Fig. \ref{fig-SpecificA1}, which stand within the finitely correlated region.

For $|J_b|>1$ and $h=0$ [the $B_1$ boundary, Fig. \ref{fig-Ground} (c)], the correlation functions $\zeta$ with $L \geq 2$ is zero even at sufficiently large $\beta$, which is coincident with previous results.
\\

\subsection{Exotic correlation patterns in A$_2$ and A$_3$ phases}

In this subsection, we discuss the spin-spin correlation functions in other two nontrivial A$_2$ and A$_3$ phases with sufficiently large $\beta$.

By introducing $\zeta_{ij} = \langle s_a s_b \rangle$ with $s_a \in P_i$ and $s_b \in P_j$ ($i,j=1,\cdots,4$), we categorize the spin-spin correlations in A$_2$ phase into ten cases ($\zeta_{11}$, $\zeta_{12}=\zeta_{21}$, $\zeta_{13}=\zeta_{31}$, $\zeta_{14}=\zeta_{41}$, $\zeta_{22}$, $\zeta_{23}=\zeta_{32}$, $\zeta_{24}=\zeta_{42}$, $\zeta_{33}$, $\zeta_{34}=\zeta_{43}$, $\zeta_{44}$) and discuss them separately in the following.
\begin{figure}[tbp]
\resizebox{0.45\textwidth}{!}{%
  \includegraphics{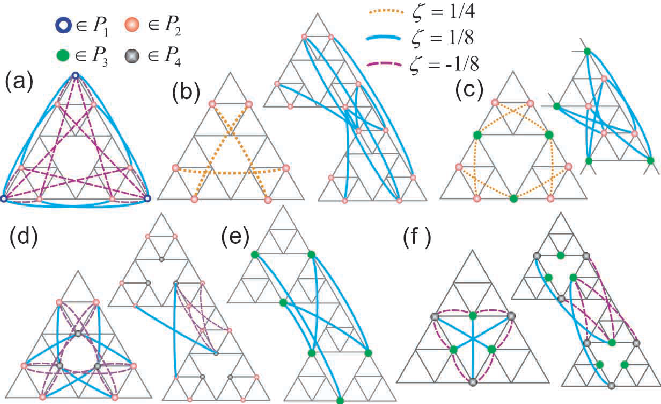}
}
\caption{(Color online) The spin-spin correlation $\langle s_a s_b \rangle$ in A$_2$ phase within two adjacent $\Delta^{(3)}$'s at sufficiently large $\beta$ with (a) $s_a \in P_1$ and $s_b \in P_2$, (b) $s_a, s_b \in P_2$, (c) $s_a \in P_2$ and $s_b \in P_3$, (d) $s_a \in P_2$ and $s_b \in P_4$, (e) $s_a, s_b \in P_3$, (f) $s_a \in P_3$ and $s_b \in P_4$. The spin $s_a$ and $s_b$ are connected by a yellow dot line with $\langle s_a s_b \rangle = 1/4$, by a blue solid line with $\langle s_a s_b \rangle = 1/8$, by a purple dash line with $\langle s_a s_b \rangle = -1/8$ and unconnected with $\langle s_a s_b \rangle = 0$.}
\label{fig-CRLPlot}
\end{figure}

(1) $\zeta_{11}=0$ for any choice of two spins that belong to the subset P$_1$.

(2) $|\zeta_{12}|=1/8$ when $s_a \in P_1$ and $s_b \in P_2$ in the same $\Delta^{(3)}$, or $\zeta_{12}=0$ otherwise. See Fig. \ref{fig-CRLPlot} (a).

(3) $\zeta_{13}=-1/8$ when $s_a \in P_1$ and $s_b \in P_3$ in the same $\Delta^{(2)}$, $\zeta_{13}=1/8$ when $s_a$ and $s_b$ in two different but adjacent $\Delta^{(2)}$'s, or $\zeta_{13}=0$ otherwise.

(4) $\zeta_{14}=1/4$ when $s_a \in P_1$ and $s_b \in P_4$ in the same $\Delta^{(2)}$ or $\zeta_{14}=0$ otherwise.

(5) It is shown in Fig. \ref{fig-CRLPlot} (b) that $\zeta_{22}$ is equal to either $1/4$ or $0$ when $s_a,s_b \in P_2$ are in the same $\Delta^{(3)}$, and $\zeta_{22}$ is equal to either $1/8$ or $0$ when $s_a$ and $s_b$ are in two different but adjacent $\Delta^{(3)}$'s. $\zeta_{22}=1/16$ for any choice of $s_a$ and $s_b$ that are in two different $\Delta^{(3)}$'s which are not adjacent. Thus for the long distance, we have the correlation $\zeta_{22}=1/16$ which is induced by the magnetic field.

(6) The pattern of $\zeta_{23}$ within one $\Delta^{(3)}$ or two adjacent $\Delta^{(3)}$'s is similar to that of $\zeta_{22}$ as shown in Fig. \ref{fig-CRLPlot} (c). For the long distance, we have $\zeta_{23}=1/16$ when $s_a$ and $s_b$ are in two non-adjacent $\Delta^{(3)}$'s, which is the same as $\zeta_{22}$.

(7) As shown in Fig. \ref{fig-CRLPlot} (d), $|\zeta_{24}|$ is equal to either $1/8$ or $0$ when $s_a \in P_2$, $s_b \in P_4$ are in the same $\Delta^{(3)}$ or in two different but adjacent $\Delta^{(3)}$. $\zeta_{24}=0$ when $s_a$ and $s_b$ are in two non-adjacent $\Delta^{(3)}$'s.

(8) $\zeta_{33}=0$ when $s_a,s_b \in P_3$ are in the same $\Delta^{(3)}$, and $\zeta_{33}=0$ or $1/8$ when $s_a$ and $s_b$ are in two adjacent $\Delta^{(3)}$'s, as shown in Fig. \ref{fig-CRLPlot} (e), and $\zeta_{33}=1/16$ when $s_a$ and $s_b$ are in two non-adjacent $\Delta^{(3)}$'s.

(9) The pattern of $\zeta_{34}$ within one $\Delta^{(3)}$ or two adjacent $\Delta^{(3)}$'s is similar to that of $\zeta_{24}$ as shown in Fig. \ref{fig-CRLPlot} (f). For a longer distance, $\zeta_{34}=0$, which is the same as $\zeta_{24}$.

(10) $\zeta_{44}=1/4$ only when $s_a,s_b \in P_4$ are in two different but adjacent $\Delta^{(2)}$'s and also in two different but adjacent $\Delta^{(3)}$'s, or otherwise $\zeta_{44}=0$ (if $s_a$ and $s_b$ are in two different $\Delta^{(2)}$'s but these two $\Delta^{(2)}$'s belong to the same $\Delta^{(3)}$, $\zeta_{44}=0$).

In A$_3$ phase, the spins belonging to the subset P$_1$ and P$_3$ are polarized, so any two spins $s_a$ and $s_b$ with $s_a, s_b \in P_1 \cup P_3$ are maximally correlated with $\zeta = 1/4$. We also calculated the correlation associated with the spins in P$_2$ and P$_4$. The results show that with $s_a \in P_1 \cup P_3$ and $s_b \in P_2 \cup P_4$ we have $\zeta = 1/12$, and with $s_a,s_b \in P_2 \cup P_4$ we have $\zeta = -1/12$ when $s_a$, $s_b$ are the nearest neighbors, and $\zeta = 1/36$ when these two spins are not the nearest neighbors.

Despite the complexity, the correlation patterns share a common feature: for any $t$, the pattern of the spin-spin correlations of $s_a$ and $s_b$ that belong to the $\Delta^{(t)}$ bears the symmetry as same as $\Delta^{(t)}$ in the SG. Specifically speaking, both the SG $\Delta^{(t)}$ and the corresponding correlation pattern are invariant under the rotations with the angles $\theta = 0, 2\pi/3, 4\pi/3, 2\pi$ around the center of $\Delta^{(t)}$ as the rotating axis. \bigskip

\section{Generalization to quantum system}

\begin{figure}[tbp]
  \resizebox{0.45\textwidth}{!}{%
  \includegraphics{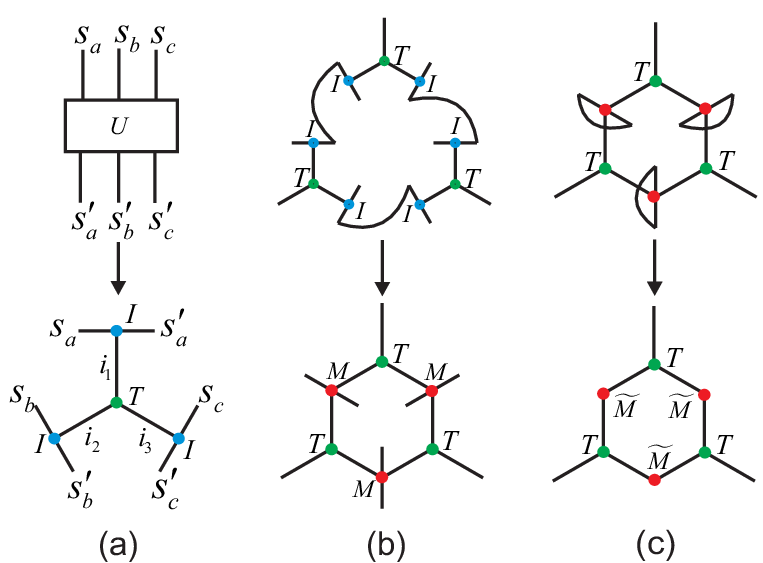}
}
  \caption{(Color online) (a) The local imaginary time evolution operator $U_{s_as_bs_cs_a's_b's_c'}$ can be written as the product of a third-order tensor $T$ and three identities $I$. (b) A fractal tensor product density operator can be constructed by contracting $T$ and three identities $I$. (c) By tracing physical indexes, the partition function can be obtained by contracting a fractal tensor network formed by $T$ and $\tilde{M}$, which can be handled similarly as the one of the Ising model.}
  \label{fig-TPDO}
\end{figure}

Below, we provide a TN-based scheme to study the finite temperature properties of the fractal Sierp\'{i}nski gasket with quantum interactions and discuss if the super-exponential relation is still retained with the presence of a small quantum fluctuation. The quantum Hamiltonian reads
\begin{equation}\label{14}
  H^Q=H+\delta \sum_{\substack{ij}}(s_{i}^xs_{j}^x+s_{i}^ys_{j}^y),
\end{equation}
where the first term $H$ is the Ising Hamiltonian along the z-direction of the spins and the second term gives the nearest-neighbor interactions that introduce quantum fluctuation. The finite temperature density operator is written as
\begin{equation}
  \rho(\beta)=e^{-\beta H^Q}=e^{-\beta\sum_{\substack{i}} H^Q_{\Delta_i}},
\end{equation}
with $\beta$ the inverse temperature and $H^Q_{\Delta_i}$ the local quantum Hamiltonian of the $i$th triangle. Then we employ the Trotter-Suzuki decomposition \cite{Trotter} as
\begin{equation}
  \rho(\beta) = [ \rho(\tau) ]^\mathcal{K} =  ( e^{-\tau\sum_i H^Q_{\Delta_i}})^\mathcal{K} = ( \prod_i e^{-\tau H^Q_{\Delta_i}})^\mathcal{K} + O(\tau),
  \label{eq-Trotter}
\end{equation}
with $\tau$ a small quantity satisfying $\mathcal{K}\tau=\beta$ and $O(\tau)$ containing higher-order small terms that originate from the commutation. The density matrix $\rho(\tau) = e^{-\tau\sum_i H^Q_{\Delta_i}}$ can be transformed into a TN form as followed. Introduce the local imaginary time evolution operator
\begin{equation}
  U_{\Delta_i}= e^{-\tau H^Q_{\Delta_i}} = \sum_{\substack{s_as_bs_cs_a's_b's_c'}}U_{s_as_bs_cs_a's_b's_c'} |s_as_bs_c\rangle\langle s_a's_b's_c'|,
\end{equation}
where $U_{s_as_bs_cs_a's_b's_c'}$ is the coefficient matrix and $|\ast \rangle$ ($\langle \ast|$) denotes the \textit{bra} (\textit{ket}) basis of a spin. The coefficient matrix is identically transformed into the product of a third-order tensor $T$ and three identities $I$ [Fig. \ref{fig-TPDO} (a)] as
\begin{equation}
U_{s_as_bs_cs_a's_b's_c'}=\sum_{\substack{i_1i_2i_3}}T_{i_1i_2i_3} I_{i_1,s_as_a'}I_{i_2,s_bs_b'}I_{i_3,s_cs_c'},
\end{equation}
Then $\rho(\tau)$ is represented by a fractal TN form dubbed as fractal tensor product density operator (fTPDO) \cite{ODTNS} which reads
\begin{equation}
\rho(\tau)= Tr [\prod_{i\in node} T_{i_1i_2i_3}^{(i)} \prod_{n\in bond} M^{(n)}_{i_ni_n'ss'}]
\end{equation}
with $M_{i_1i_2ss'}=\sum_{\substack{s''}}I_{i_1,ss''}I_{i_2,s''s'}$ and $Tr$ meaning to trace all shared indexes. With the TN representation of $\rho(\tau)$, the finite temperature density matrix $\rho(\beta)$ can be achieved by the imaginary time evolution algorithm \cite{iTEBD}.

Now we consider a simpler case with $|\delta| \ll 1$, implying that the quantum fluctuation terms can be treated as a perturbation. Then $\rho(\beta)$ for any $\beta$ can be straightforwardly written in the fTPDO form by following exactly the same scheme that is given above. After summing over physical indexes by $\tilde{M}^{(n)}_{i_ni_n'} = \sum_s M^{(n)}_{i_ni_n'ss}$ [Fig. \ref{fig-TPDO} (b)], the partition function $Z(\beta)$ becomes the contraction of a fractal TN as
\begin{equation}
Z(\beta)= Tr [\prod_{i\in node} T_{i_1i_2i_3}^{(i)} \prod_{n\in bond} \tilde{M}^{(n)}_{i_ni_n'}],
\end{equation}
which can be handled in the same way as the Ising case.

\begin{figure}[tbp]
\resizebox{0.45\textwidth}{!}{%
  \includegraphics{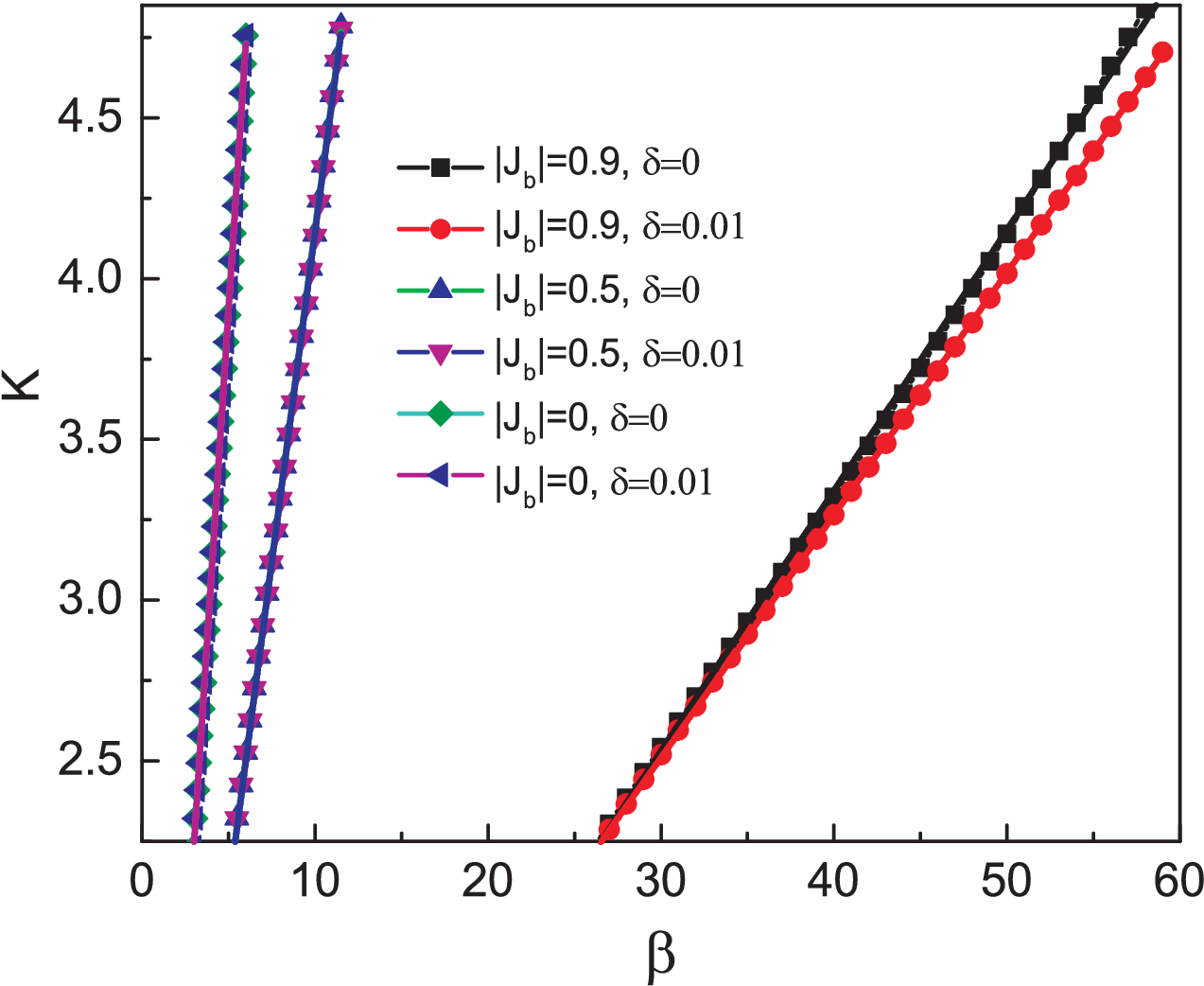}
}
\caption{The $\beta$ dependence of $K=\ln[4\ln(\xi)/\ln2]$ with quantum fluctuation. The regression coefficient of each linear fitting is larger than 0.999.}
\label{fig-QF}
\end{figure}

It is interesting to see whether the super-exponential relation that is observed in the Ising case will be destroyed by a small quantum fluctuation. One can observe from Fig. \ref{fig-QF} that for different $|J_b|$ and $\delta$, the relation of Eq.(\ref{eq-xi}) still holds, though the correlation length is decreased by the quantum fluctuation especially for $|J_b|\simeq1$.

\section{Summary}

In summary, we have systematically studied the anisotropic antiferromagnetic Ising model on the fractal Sierpi\'{n}ski gasket where exotic properties due to both the classical frustration and the special geometry of the Sierpi\'{n}ski gasket are disclosed. The zero temperature phase diagram with four phases is presented, and the thermodynamic properties of the three nontrivial phases are explored by calculating the magnetization, residual entropy, specific heat and magnetic susceptibility. The spin-spin correlations are also probed. For $h=0$, the thermodynamic crossover behavior with $0 \leq |J_b|<1$ and the zero-temperature crossover from $0 \leq |J_b|<1$ to $|J_b|>1$ are both analyzed by studying the temperature dependence of the correlation length. The exotic correlation patterns are disclosed in the $1/3$ magnetization plateau disordered phase and the $5/9$ plateau partially ordered ferrimagnetic phase. Furthermore, we generalized our scheme to study the fractal Sierpi\'{n}ski gasket with Heisenberg interactions, with which we discover that the super-exponential relation of the correlation length holds after introducing a quantum fluctuation.\\

\section*{Author contribution statement}

Shi-Ju Ran and Meng Wang made a major contribution to this paper. All other authors designed the research and write the manuscript equally.

The authors are indebted to W. Li, Z. C. Wang, X. Yan, Z. G. Zhu and C. J. Chen for useful discussions. This work is supported in part by the NSFC (Grants No. 90922033 and No. 10934008), the MOST of China (Grant No. 2012CB932900 and No. 2013CB93
3401), and the CAS.


\begin{thebibliography}{99}

\bibitem{Potts} F. Y. Wu, Rev. Mod. Phys. \textbf{54}, 235 (1982).

\bibitem{Frustrate} R. Moessner and A. R. Ramirez, Physics Today \textbf{59}, 24 (2006).

\bibitem{ReEntro} R. Shrock and S. H. Tsai, Phys. Rev. E \textbf{56}, 4111 (1997).

\bibitem{PO} P. Azaria, H. T. Diep, and H. Giacomini, Phys. Rev. Lett. \textbf{59}, 1629 (1987).

\bibitem{liw} W. Li, S. S. Gong, Y. Zhao, S. J. Ran, S. Gao, and G. Su, Phys. Rev. B \textbf{82}, 134434 (2010).

\bibitem{POIrregular} Q. N. Chen, M. P. Qin, J. Chen, Z. C. Wei, H. H. Zhao, B. Normand, and T. Xiang, Phys. Rev. Lett. \textbf{107}, 165701 (2011).

\bibitem{zhaoy} Y. Zhao, W. Li, B. Xi, Z. Zhang, X. Yan, S.J. Ran, T. Liu, and G. Su, Phys. Rev. E \textbf{87}, 032151 (2013).

\bibitem{Fractal1} Y. Gefen, B. B. Mandelbrot, and A. Aharony, Phys. Rev. Lett. \textbf{45},855 (1980).

\bibitem{Fractal11} Y. Gefen, A. Aharony, B. B. Mandelbrot and S. Kirkpatrick, Phys. Rev. Lett. \textbf{47}, 1771 (1981).

\bibitem{Fractal12} Y. Gefen, A. Aharony, Y. Shapir and B. B. Mandelbrot, J. Phys. A \textbf{17}, 435 (1984).

\bibitem{Fractal2} S. H. Liu, Phys. Rev. B \textbf{32},5804 (1985).

\bibitem{Fractal3} T. Sto\v{s}i\'{c}, B. Sto\v{s}i\'{c}, S. Milo\v{s}evi\'{c}, and H. E. Stanley, Phys. Rev. E \textbf{49},1009 (1994); B. Kutnjak-Urbanc, S. Zapperi, S. Milo\v{s}evi\'{c}, and H. E. Stanley, Phys. Rev. E \textbf{54},272 (1996).

\bibitem{Fractal4} A. Voigt, W. Wenzel, J. Richter, and P. Tomczak, Eur. Phys. J. B \textbf{38}, 49 (2004).

\bibitem{Fractal5} S. C. Chang and R. Shrock, Phys. Lett. A \textbf{377}, 671 (2013).

\bibitem{Fractal6} L. Tian, H. Ma, W. A. Guo, and L. H. Tang, Eur. Phys. J. B \textbf{86}, 197 (2013).

\bibitem{Material1}M.Gonzalez, F. Cervantes-Lee, and L. W. ter Haar, Mol. Cryst. Liq. Cryst. \textbf{233}, 317(1993).

\bibitem{Material2}S. Maruti and L. W. ter Haar, J. Appl. Phys. \textbf{75}, 5949(1994).

\bibitem{Material3}S.Ateca, S. Maruti and L. W. ter Haar, J. Magn. Magn. Mater.  \textbf{147}, 398(1995).

\bibitem{Material4}M. Mekata, M. Abdulla, T. Asano, H. Kikuchi, T. Goto, T. Morishita, and H.Hori, J. Magn. Magn. Mater.  \textbf{177-181}, 731(1998).

\bibitem{Material5}S. Okubo, M. Hayashi, S. Kimura, H. Ohta, M. Motokawa, H. Kikuchi, and H. Nagasawa, Physica B  \textbf{246-247}, 553(1998).

\bibitem{Material6}M. Mekata, M. Abdulla, M. Kubota, and Y. Oohara, Can. J. Phys.  \textbf{79}, 1409(2001).

\bibitem{Lattice1}J. Stre\v{c}ka, L. \v{C}anov\'{a}, M. Ja\v{s}\v{c}ur, M.Hagiwara, Phys. Rev. B  \textbf{78}, 024427(2008).

\bibitem{Lattice2}D. X. Yao, Y. L. Loh, E. W. Carlson, and M. Ma, Phys. Rev. B  \textbf{78}, 024428(2008).

\bibitem{Lattice3}J. Stre\v{c}ka and L. \v{C}anov\'{a}, J.Phys.: Conf. Ser. \textbf{145}, 012012(2009).

\bibitem{Lattice4}J. Stre\v{c}ka, J. Magn. Magn. Mater.\textbf{316}, e346(2007).

\bibitem{Lattice5}M. Isoda, H. Nakano, and T. Sakai, J. Phys.:Conf. Ser. \textbf{320}, 012010(2011).

\bibitem{Lattice6}M. Isoda, H. Nakano, and T. Sakai, J. Phys. Soc. Jpn. \textbf{81}, 053703(2012).

\bibitem{Lattice7}Y.-H. Chen, H.-S. Tao,D.-X. Yao, and W.-M. Liu, Phys. Rev. Lett. \textbf{108}, 246402(2012).

\bibitem{TIT1}J. \v{C}is\'{a}rov\'{a} and J. Stre\v{c}ka, Phys. Rev. B  \textbf{87}, 024421(2013).

\bibitem{TIT2}J. \v{C}is\'{a}\'{a}, F. Michaud, F. Mila, and J. Stre\v{c}ka, Phys. Rev. B  \textbf{87}, 054419(2013).

\bibitem{TIT3}J. Stre\v{c}ka and C. Ekiz, Phys. Rev. E  \textbf{91}, 052143(2015).

\bibitem{Tc} See R. B. Griffiths, et al., \textit{in Phase Transitions and Critical Phenomena} (Academic Press, New York, 1972).

\bibitem{Residual} T. Sto\v{s}i\'{c}, B. Sto\v{s}i\'{c}, S. Milo\v{s}evi\'{c}, and H. E. Stanley, Phys. Rev. A. \textbf{37}, 1747 (1988).

\bibitem{TRG} M. Levin and C. P. Nave, Phys. Rev. Lett. \textbf{99}, 120601 (2007).

\bibitem{Recursive} H. Moraal, \textit{Classical, Discrete Spin Models: Symmetry, Duality and Renormalization} (Springer-Verlag, Berlin Heidelberg, 1984) p.208.

\bibitem{RGT} In all of our calculations, we assume $\beta=200$ as a sufficiently low temperature for ground state properties, and take the recursive step $t=10^3$, where the total number of spins is determined by $N^{(t)}=(3+3^t)/2$.

\bibitem{Trotter} M. Suzuki and M. Inoue, Prog. Theor. Phys. \textbf{78}, 787 (1987); M. Inoue and M. Suzuki, Prog. Theor. Phys. \textbf{79}, 645 (1988).

\bibitem{iTEBD} G. Vidal, Phys. Rev. Lett. \textbf{91}, 147902 (2003); Phys. Rev. Lett. \textbf{98},070201 (2007).

\bibitem{ODTNS} S. J. Ran, W. Li, B. Xi, Z. Zhang, and G. Su, Phys. Rev. B \textbf{86},  134429 (2012).
%
\end{thebibliography}
\end{document}